\documentclass[journal,10pt]{IEEEtran} 

\pdfoutput=1

\makeatletter
\def\ps@headings{%
\def\@oddhead{\mbox{}\scriptsize\rightmark \hfil \thepage}%
\def\@evenhead{\scriptsize\thepage \hfil \leftmark\mbox{}}%
\def\@oddfoot{}%
\def\@evenfoot{}}
\makeatother
\usepackage{color}
\usepackage[cmex10]{amsmath}
\usepackage{comment}
\usepackage[pdftex]{graphicx}
\usepackage{bm}
\usepackage{amsthm}
\usepackage{amssymb}
\usepackage{enumerate}
\usepackage{stfloats}
\usepackage{cases}
\usepackage{epstopdf}
\usepackage{thmtools}
\usepackage{graphicx}
\usepackage{mathtools}
\usepackage{soul} 
\usepackage{hyperref}
\usepackage{pgfplots}
\usepackage{etoolbox}
\usepackage{makecell}
\usepackage{algorithm}
\usepackage{algorithmic}
\usepackage{tikz}
\usepackage{verbatim}
\usepackage{comment}
\usetikzlibrary{shapes.geometric, arrows, positioning, matrix, patterns}
\usepackage{nccmath}
\usepackage{float}
\usepackage{graphicx}

\theoremstyle{definition}

\AfterEndEnvironment{problem}{\noindent\ignorespaces}

\AfterEndEnvironment{lemma}{\noindent\ignorespaces}
\newtheorem {definition}{Definition}
\AfterEndEnvironment{definition}{\noindent\ignorespaces}

\AfterEndEnvironment{theorem}{\noindent\ignorespaces}
\newtheorem {proposition}{Proposition}
\AfterEndEnvironment{proposition}{\noindent\ignorespaces}

\begin{document}

\title{\huge{A Distributed Power Control Algorithm for Energy Efficiency Maximization in Wireless Cellular Networks}}

\author{Rojin~Aslani, \emph{Student Member, IEEE}, and Mehdi~Rasti, \emph{Member, IEEE}
\thanks{The authors are with the Department of Computer Engineering, Amirkabir University of Technology, Tehran 1591634311, Iran (e-mail: rojinaslani@aut.ac.ir; rasti@aut.ac.ir).}
\thanks{Digital Object Identifier 10.1109/LWC.2020.3010156}}

\makeatletter
\def\ps@IEEEtitlepagestyle{
  \def\@oddfoot{\mycopyrightnotice}
  \def\@evenfoot{}
}
\def\mycopyrightnotice{
  {\footnotesize
  \begin{minipage}{\textwidth}
  \centering
  2162-2345~\copyright2020 IEEE. 
Personal use is permitted, but republication/redistribution requires IEEE permission. \\
 See https://www.ieee.org/publications/rights/index.html for more information.
  \end{minipage}
  }
}

\maketitle

\begin{abstract}
In this paper, we propose a distributed power control algorithm for addressing the global energy efficiency (GEE) maximization problem subject to satisfying a minimum target SINR for all user equipments (UEs) in wireless cellular networks. We state the problem as a multi-objective optimization problem which targets minimizing total power consumption and maximizing total throughput, simultaneously, while a minimum target SINR is guaranteed for all UEs. 
We propose an iterative scheme executed in the UEs to control their transmit power using individual channel state information (CSI) such that the GEE is maximized in a distributed manner. We prove the convergence of the proposed iterative algorithm to its corresponding unique fixed point also shown by our numerical results. Additionally, simulation results demonstrate that our proposed scheme outperforms other algorithms in the literature and performs like the centralized algorithm executed in the base station and maximizes the GEE using the global CSI.  
\end{abstract}

\begin{IEEEkeywords}
Distributed power control; energy efficiency; multi-objective optimization problem; wireless cellular networks.

\end{IEEEkeywords}

\IEEEpeerreviewmaketitle

\section{Introduction}
Nowadays, energy efficiency (EE) is a leading concern in wireless cellular networks. 
In \cite{Zappone}, three metrics for EE are defined: 1) Minimum-EE defined as the minimum EE of user equipments (UEs) in which the EE of a UE described by the ratio of the UE's throughput and its consumed power; 2) Sum-EE defined as the summation over the EE of UEs; 3) Global EE (GEE) defined as the ratio between the total throughput and the total power consumption of the system. As the GEE is the most commonly used metric for EE maximization problems in wireless cellular networks \cite{Aslani2019}-\cite{khalili}, we focus on it in this paper. 

To maximize the GEE, it is required that the total throughput is maximized while the minimum power is consumed. However, the objectives of total throughput maximization and power consumption minimization are contradicting each other. Since, for total throughput maximization, the UEs require to transmit with high power while high power increases the total power consumption. Thus, the transmit power control has a significant role in maximizing the GEE in cellular networks.

There is a lot of research that focuses on designing power control schemes to maximize the GEE in cellular networks \cite{Zappone}-\cite{Dong}.
In \cite{Zappone}, two centralized power control algorithms are proposed to address the problems of the GEE maximization and the weighted minimum-EE maximization using fractional programming and sequential convex optimization framework. 
Employing the fractional programming and Dinkelbach algorithm, a joint power control scheme for the uplink and downlink of a cellular network is proposed in \cite{Aslani2019} which aims at maximizing the GEE while the quality of service requirements of the UEs are guaranteed. 
In \cite{Tang} and \cite{Ng}, the problem of the GEE maximization in the downlink of cellular networks is studied.
The authors of \cite{khalili} and \cite{Amin} apply the multi-objective optimization (MOO) approach to address the problem of the GEE-spectral efficiency tradeoff.
In \cite{Ozek} and \cite{Liu}, power control schemes are proposed to maximize the GEE in D2D communications underlaying cellular networks.

The existing power control algorithms for the GEE maximization are mostly centralized \cite{Zappone}-\cite{Liu}. However, distributed power control schemes are practically preferred to centralized ones due to the utilization of local information and minimal feedback from the base station (BS). 
The authors of \cite{Dong} propose a semi-distributed algorithm that aims at maximizing the GEE in the downlink of a cellular network.
Employing game theory, a distributed power control scheme is presented in \cite{Zappone} to maximize the sum-EE. 
In \cite{Zhang}, a threshold-based distributed power control scheme is proposed to address the sum-EE maximization problem.
The fixed-target-SIR-tracking power control (TPC) algorithm is proposed in \cite{TPC} for the aggregate transmit power minimization while a minimum target SINR is satisfied for all UEs. 
In \cite{OPC1} and \cite{OPC2}, an opportunistic power control scheme to maximize the total throughput is proposed. 
The distributed dynamic target-SIR tracking power control (DTPC) algorithm is presented in \cite{DTPC} which maximizes the total throughput while guarantees a minimum target SINR for all UEs. 
The authors of \cite{Rasti2011} propose a distributed power control algorithm with temporary removal and feasibility check to address the gradual removal problem in wireless networks.
In \cite{Aslani}, two distributed power control schemes are presented for the aggregate transmit power minimization and the total throughput maximization in energy harvesting wireless networks.
Most of the proposed distributed power control algorithms in the literature for cellular networks focused on objectives of total throughput maximization, total power consumption minimization, and sum-EE maximization.

In this work, we propose a distributed power control scheme for GEE maximization by simultaneously maximizing the total throughput and minimizing the total power consumption while satisfying a minimum SINR for UEs. To do this, we apply MOO framework which makes a balance between the competing objectives. 
The contributions of this~work~are~as~follows:
\begin{itemize}
\item
We study the power control problem to maximize GEE subject to the minimum target SINR for the UEs in the uplink of cellular networks (in contrast to \cite{Dong} which considers downlink communications). This is in contrast with other existing literature such as \cite{Zappone}, \cite{Zhang}-\cite{Aslani}, in which the objective functions are sum-EE maximization, total throughput maximization, or total power minimization.
\item 
We propose a distributed power control scheme to maximize GEE. This is in contrast to \cite{Zappone}-\cite{Liu} which propose centralized schemes. For this purpose, since the stated problem is non-convex, we first reformulate it as a MOO problem which targets minimizing the total power consumption and maximizing total throughput, simultaneously. We then employ $\epsilon$-constraint method \cite{khalili} to address the MOO problem. Finally, we propose an iterative scheme that is distributed and executed in the UEs to control their transmit power using individual channel state information~(CSI).
\item
We provide the system feasibility condition. We prove that the proposed iterative algorithm converges to its unique fixed point also shown by the numerical results. Additionally, simulation results demonstrate that our distributed algorithm which uses the local CSI performs like the centralized algorithm proposed in \cite{Aslani2019} in which the global CSI is assumed to be presented in the BS.
\end{itemize}

\section{System Model and Problem Formulation}
We consider a single cell in a wireless CDMA network with one BS and $K$ UEs, whose set is denoted by $\mathcal{K}=\{1,\cdots,K\}$. Let $h_i$ denote the channel gain from $i$th UE to the BS. Also, $p_i$ denotes the transmit power of the $i$th UE and $0 \leq p_i \leq \overline{p}_i$, where $\overline{p}_i$ is the maximum transmit power for the $i$th UE. We assume the presence of  additive white Gaussian noise (AWGN) with power $\sigma^2$ at the BS. 

Given the transmit power vector $\bold{p}$, $\gamma_i(\bold{p})$ denotes the SINR of the $i$th UE at the BS receiver, given by:
\begin{equation}
\label{SINR}
\gamma_i(\bold{p})=\frac{p_i h_i}{I_i(\bold{p})},
\end{equation}
where, $I_i(\bold{p})$ is the total interference caused to the $i$th UE and given by $I_i(\bold{p})\!=\!\sum_{j \in \mathcal{K},j \neq i} {p_j h_j} \!+\! \sigma^2$. The effective interference for the $i$th UE, denoted by $\phi_i(\bold{p})$, is defined as: 
\begin{equation}
\label{effective interference}
\phi_i(\bold{p})=\frac{I_i(\bold{p})}{h_i}.
\end{equation}
The value of $\phi_i(\bold{p})$ shows the quality of the $i$th UE's channel, lower interference and higher channel gain lead to lower $\phi_i(\bold{p})$, implying a good channel, compared with higher interference and lower channel gain, which lead to higher $\phi_i(\bold{p})$, implying a poor channel \cite{DTPC}.
According to Shannon formula, the throughput for the $i$th UE (bps/Hz) is given by:
\begin{equation}
\label{throughput}
T_i(\bold{p}) = \log( 1+\gamma_i(\bold{p}) ).
\end{equation}
The total throughput is obtained by $T(\bold{p})=\sum_{i=1}^K{T_i(\bold{p})}$. The total power consumption is formed as:
\begin{equation}
\label{totalPowerConsumption}
P^\textnormal{T}(\bold{p}) = P_\textnormal{BS}^\textnormal{C} + \sum_{i=1}^K{P_i^\textnormal{C}} + \sum_{i=1}^K{\mu_i p_i},
\end{equation}
where, $P_\textnormal{BS}^\textnormal{C}$ and $P_i^\textnormal{C}$ are the power consumed by the circuit of the BS and $i$th UE, respectively, and $\mu_i$ is the power amplifier inefficiency of  $i$th UE \cite{Aslani2019}. We define the GEE as the ratio of total throughput and total power consumption, given by:
\begin{equation}
\label{energyEfficiency}
EE(\bold{p})=\frac{T(\bold{p})}{P^\textnormal{T}(\bold{p})}.
\end{equation}

The problem of power control for maximizing the GEE subject to the minimum target SINR requirements for UEs and the feasibility of transmit power for UEs is formulated~as: 
\begin{align}
\label{problem}
\underset{\bold{p}}{\mathrm{maximize}}& \quad EE(\bold{p})  \\
\mathrm{s.t.}~~~ 
\textnormal{C}_1: & \quad \gamma_i(\bold{p}) \geq \widehat{\gamma}_i, & \forall i \in \mathcal{K} \nonumber \\
\textnormal{C}_2: & \quad 0 \leq p_i \leq \overline{p}_i, & \forall i \in \mathcal{K} \nonumber 
\end{align}
where $\textnormal{C}_1$ corresponds to the target SINR requirement for UEs. 
The feasibility of transmit power for UEs is met by $\textnormal{C}_2$. 

The problem (\ref{problem}) is non-convex due to the fractional objective function which is an obstacle to address it in a distributed manner. In the next section, we address (\ref{problem}) by reformulating it as an equivalent MOO problem to minimize aggregate transmit power and maximize total throughput, simultaneously.

\section{The Proposed Distributed Power Control Scheme and Its Analysis}
As given in (\ref{energyEfficiency}), the GEE is the ratio of the total throughput to the total power consumption. Thus, maximizing the GEE is equivalent to minimizing the total power consumption while maximizing the total throughput, simultaneously \cite{Amin}. On the other hand, as the circuit power consumption of BS and UEs, i.e., $P_\textnormal{BS}^\textnormal{C}$ and $P_i^\textnormal{C}$, respectively, as well as the power amplifier inefficiency of UEs, i.e., $\mu_i$, are fixed, we~can~conclude~that minimizing the total power consumption is~equivalent~to~minimizing the aggregate transmit power. Accordingly,~we~reformulate (\ref{problem}) as an equivalent MOO problem,~that~is:
\begin{align}
\label{multi-objective problem}
\underset{~~~~\bold{p}}{f_1: \mathrm{minimize}}& \quad \sum_{i=1}^K{p_i}  \\
\underset{~~~~\bold{p}}{f_2: \mathrm{maximize}}& \quad T(\bold{p}) \nonumber \\
\mathrm{s.t.}\qquad~~ & \quad \textnormal{C}_1-\textnormal{C}_2, \nonumber 
\end{align}
where, $f_1$ is the aggregate transmit power minimization and $f_2$ is the total throughput maximization.

\begin{proposition}
The MOO problem in (\ref{multi-objective problem}) is equivalent to the GEE maximization problem in (\ref{problem}).
\end{proposition}
\vspace{-20pt}
\begin{proof}
Suppose $\mathcal{D}$ is the set of feasible solutions to (\ref{problem}) spanned by constraints $\textnormal{C}_1-\textnormal{C}_2$. We denote $q^*$ as the optimal GEE in (\ref{problem}) represented by $q^* \!=\! \frac{T(\bold{p}^*)}{P^\textnormal{T}(\bold{p}^*)} \!=\! \underset{\bold{p} \in \mathcal{D}}{\mathrm{maximize}}{ \frac{T(\bold{p})}{P^\textnormal{T}(\bold{p})} }$, where $\bold{p}^*$ is the optimal transmit power vector for (\ref{problem}). It was proved in \cite{Dinkelbach} that for the fractional objective function in (\ref{problem}), there is an equivalent subtractive form as follows:
\begin{equation}
\label{equivalent problem}
\underset{\bold{p} \in \mathcal{D}}{\mathrm{maximize}} \quad  T(\bold{p})-q^*P^\textnormal{T}(\bold{p}),
\end{equation}
which shares the same optimal solution, i.e., $\bold{p}^*$. The objective function in (\ref{equivalent problem}) represents a linear scalarization for a multi-objective function in the form of $\underset{\bold{p} \in \mathcal{D}}{\mathrm{maximize}} ~ \{T(\bold{p}), -P^\textnormal{T}(\bold{p})\}$ with the weights of $1$ and $q^*$, respectively. The objective of $\underset{\bold{p} \in \mathcal{D}}{\mathrm{maximize}} ~ -P^\textnormal{T}(\bold{p})$ can be rewritten as $\underset{\bold{p} \in \mathcal{D}}{\mathrm{minimize}} ~ P^\textnormal{T}(\bold{p})$. Considering the variable part of total power consumption $P^\textnormal{T}(\bold{p})$, the problem (\ref{equivalent problem}) can be rewritten as the problem (\ref{multi-objective problem}). This completes the proof.
\end{proof}

Now, we find a solution for MOO problem (\ref{multi-objective problem}) by employing $\epsilon$-constraint method \cite{khalili}. To do this, we keep $f_1$ as the primary objective function and redefine $f_2$ as a constraint. Accordingly, the new problem is stated as:
\begin{align}
\label{new problem}
\underset{~~~~\bold{p}}{f_1: \mathrm{minimize}}& \quad \sum_{i=1}^K{p_i}  \\
\mathrm{s.t.}~~
\textnormal{C}_0: & \quad T(\bold{p}) \geq \epsilon  \nonumber \\
& \!\!\!\!\!\!\!\!\!\! \textnormal{C}_1-\textnormal{C}_2, \nonumber 
\end{align}
where, $\textnormal{C}_0$ ensures that the total throughput is greater than $\epsilon$. The feasibility of (\ref{new problem}) and the closeness between its solution and the solution of (\ref{problem}) highly depend on the value of $\epsilon$. To analyze the significance of the parameter $\epsilon$ on the solution of (\ref{new problem}), we consider three following cases in a similar way to \cite{khalili}:
\begin{enumerate}[~~]
\item \textbf{Case 1:} If $\epsilon=0$, then (\ref{new problem}) would turn into the problem of the aggregate transmit power minimization subject to the minimum target SINR constraints for all UEs which is a conventional power control problem in cellular networks and addressed by the TPC algorithm proposed in \cite{TPC}.
\item \textbf{Case 2:} If $\epsilon=T^\textnormal{max}$, where $T^\textnormal{max}$ is the maximum total throughput, then (\ref{new problem}) would turn into the problem of the total throghput maximization subject to the minimum target SINR constraints for all UEs which is a conventional power control problem in cellular networks and addressed by the DTPC algorithm proposed in \cite{DTPC}. 
\item \textbf{Case 3:} If $\epsilon \geq T^\textnormal{max}$, then (\ref{new problem}) would be infeasible. 
\end{enumerate}
With this analysis, we conclude that the solution of (\ref{new problem}) highly depends on the value of $\epsilon$. More specifically, the parameter $\epsilon$ makes a trade-off between the total power consumption and the total throughput. Therefore, we require to find a value for $\epsilon$ corresponding to the maximum ratio of the total throughput and the total power consumption, i.e., the GEE. 

From the above cases, it is derived that the minimum value of $\epsilon$ is $0$ and the maximum value that $\epsilon$ can take without making the problem (\ref{new problem}) infeasible is the value of $T^\textnormal{max}$ obtained by applying the DTPC algorithm. Thus, we have $0 \!\leq\! \epsilon \!\leq\! T^\textnormal{max}$. Let define $\epsilon$ as $\epsilon=\delta T^\textnormal{max}$, where $\delta \in [0,1]$. The value of the GEE modifies depends on the value of $\delta$; however, the maximum GEE is achieved for a specific value of $\delta$. 
To find the value of $\delta$, by replacing $T(\bold{p})$ and $\epsilon$, we rewrite the constraint $\textnormal{C}_0$ in problem (\ref{new problem}) as $\sum_{i=1}^{K}{T_i(\bold{p})} \geq \sum_{i=1}^{K}{\delta_i} T^\textnormal{max}$, where, $\delta_i \in [0,1]$ and $\sum_{i=1}^{K}{\delta_i}=\delta$. Indeed, $\delta_i T^\textnormal{max}$ is the throughput portion of $\delta T^\textnormal{max}$ for the $i$th UE. Thus, we can rewrite $\textnormal{C}_0$ as:
\begin{equation}
\label{C0 seperatly}
\textnormal{C}_0: T_i(\bold{p}) \geq \delta_i T^\textnormal{max}, \qquad \quad \forall i \in \mathcal{K}.
\end{equation}
The problem of finding $\delta_i$ is formulated as:
\begin{align}
\label{problem delta}
\underset{\bold{p}, \delta_i}{\mathrm{maximize}}& \quad  EE(\bold{p})  \\
\mathrm{s.t.}~~
\textnormal{C}_0: & \quad T_i(\bold{p}) \geq \delta_i T^\textnormal{max}, & \forall i \in \mathcal{K}  \nonumber \\
& \!\!\!\!\!\!\!\!\!\! \textnormal{C}_1-\textnormal{C}_2, \nonumber \\
& \!\!\!\!\!\!\!\!\!\! \textnormal{C}_3:~~ 0 \leq \delta_i \leq 1, & \forall i \in \mathcal{K}. \nonumber
\end{align}
The problem (\ref{problem delta}) can be addressed by applying centralized schemes such as the proposed algorithm in \cite{Aslani2019}.

Now, given $\delta_i$ obtained as mentioned above, we propose an iterative algorithm to address (\ref{new problem}). To do this, by replacing $T_i(\bold{p})$ from (\ref{throughput}), we rewrite $\textnormal{C}_0$ in (\ref{C0 seperatly}) as:
\begin{equation}
\label{C0 new}
\textnormal{C}_0: \gamma_i(\bold{p}) \geq 2^{\delta_i T^\textnormal{max}}-1, \qquad \quad \forall i \in \mathcal{K}.
\end{equation}
On the other hand, we have $\gamma_i(\bold{p}) \geq \widehat{\gamma}_i,~~\forall i \in \mathcal{K}$ in $\textnormal{C}_1$. Combining constraints $\textnormal{C}_0$ and $\textnormal{C}_1$, we rewrite them in a new constraints denoted by $\textnormal{C}^\prime_1$ and defined as:
\begin{equation}
\label{C1 prime}
\textnormal{C}^\prime_1: \gamma_i(\bold{p}) \geq \lambda_i, \qquad \quad \forall i \in \mathcal{K},
\end{equation}
where, $\lambda_i=\max \big\{ \widehat{\gamma}_i,2^{\delta_i T^\textnormal{max}}-1 \big\}$. Thus, (\ref{new problem}) is rewritten as:
\begin{align}
\label{final problem}
\underset{\bold{p}}{\mathrm{minimize}}& \quad \sum_{i=1}^K{p_i}  \\
& \!\!\!\!\!\!\!\!\!\!\!\!\!\!\!\!\!\!\!\!\!\!\! \mathrm{s.t.}~~
\textnormal{C}^\prime_1-\textnormal{C}_2. \nonumber 
\end{align}
Inspired by the TPC algorithm, we propose an iterative distributed transmit power control algorithm solving (\ref{final problem}). Accordingly, each UE $i$~updates~its~transmit power at iteration $t$ based on the following power updating function:
\begin{equation}
\label{power updating function}
p_i(t\!+\!1)\!=\!f_i(\bold{p}(t))\!=\!\min \Big\{ \overline{p}_i, \lambda_i \phi_i(\bold{p}(t)) \Big\}.
\end{equation}

The proposed scheme executed in each UE $i$ is given in Algorithm \ref{algorithm}. Note that in each iteration, each UE $i$ updates its transmit power based on the local information and individual CSI\footnote{The value of $\delta_i$ can be calculated at the BS and sent to the UEs.}, so the proposed algorithm is distributed. 
More specifically, by rearranging (\ref{SINR}), we have $\displaystyle{h_i=\frac{\gamma_i(\bold{p})I_i(\bold{p})}{p_i}}$. Now, by replacing $h_i$ in (\ref{effective interference}), we rewrite it as:
\begin{equation}
\label{effective interference new}
\phi_i(\bold{p})=\frac{p_i}{\gamma_i(\bold{p})}.
\end{equation}
Thus, to obtain $\phi_i(\bold{p}(t))$, UE $i$ needs only $p_i(t)$ and $\gamma_i(\bold{p}(t))$.

\begin{algorithm}
\caption{Iterative Distributed Power Control Algorithm}
\label{algorithm}
\centering
\begin{algorithmic}
\REQUIRE{$\overline{p}_i$, $\widehat{\gamma}_i$, $\delta_i$, $h_i$, $t=0$, $p_i(0)$} \\
\REPEAT
\STATE {Receive the SINR $\gamma_i(\bold{p}(t))$} \\
\STATE {Calculate $\phi_i(\bold{p}(t))$ using (\ref{effective interference new})} \\
\STATE {Update $p_i(t+1)$ using (\ref{power updating function})} \\
\STATE {Set $t=t+1$} \\
\UNTIL {Convergence} \\
\RETURN {$p_i(t)$}
\end{algorithmic}
\end{algorithm}

\subsection{Convergence Analysis}
In this section, we prove that the distributed power control scheme in (\ref{power updating function}) converges to its unique fixed point by applying two-sided scalable framework \cite{OPC2} presented in Definition~\ref{two sided scalability definition}.
\begin{definition}
\label{two sided scalability definition}
A power updating function $\medmath{\bold{f}(\bold{p})}=\medmath{[f_1(\bold{p}),f_2(\bold{p}),\!\cdots\!,f_K(\bold{p})]^T}$ is two-sided scalable if $\forall \medmath{a\!>\!1}$, $\medmath{(1/a)\bold{p} \leq \bold{p'} \leq a\bold{p}}$ implies $\medmath{(1/a)f_i(\bold{p}) \leq f_i(\bold{p'}) \leq af_i(\bold{p})}$ \cite{OPC2}.
\end{definition}

\begin{proposition}
\label{two sided scalability theorem}
The transmit power updating function $\bold{f}(\bold{p})$ in (\ref{power updating function}) is two-sided scalable.
\end{proposition}
\vspace{-20pt}
\begin{proof}
Given $a>1$, from $(1/a)\bold{p} \!\leq\! \bold{p'} \!\leq\! a\bold{p}$, we have $(1/a) \lambda_i \phi_i(\bold{p}) \!\leq\! \lambda_i \phi_i(\bold{p'}) \!\leq\! a \lambda_i \phi_i(\bold{p})$.
This implies $(1/a)f_i(\bold{p}) \!\leq\! f_i(\bold{p'}) \!\leq\! af_i(\bold{p})$. This completes the proof.
\end{proof}

\begin{proposition}
\label{fixed point lemma}
The transmit power updating function $\bold{f}(\bold{p})$ in (\ref{power updating function}) has a unique fixed point and the proposed power control algorithm $\bold{p}(t+1)=\bold{f}(\bold{p}(t))$ converges to it.
\end{proposition}
\vspace{-20pt}
\begin{proof}
In \cite{OPC2}, it is proved that for a given two-sided scalable function $\bold{f}(\bold{p})$, if there is $\bold{{a, b}}\!>\!0$ such that $\medmath{\bold{a} \!\leq\! \bold{f}(\bold{p}) \!\leq\! \bold{b}}$, then a unique fixed point exists which the power updating function $\bold{p}(t+1)=\bold{f}(\bold{p}(t))$ converges to. For the transmit power updating function $\bold{f}(\bold{p})$ in (\ref{power updating function}), since $0 \!<\! f_i(\bold{p}) \!\leq\! \overline{p}_i, ~ \forall i \in \mathcal{K}$, then by considering $\bold{a}=[a_1,\!\cdots\!, a_K]$ where $a_i\!=\!q^{\textnormal{min}}, \forall i \in \mathcal{K},$ in which $q^{\textnormal{min}} \!\rightarrow\! 0^+$ and $\bold{b}=[b_1,\!\cdots\!,b_K]$ where $b_i=\overline{p}_i,$ $~ \forall i \in \mathcal{K}$, we have $\bold{a} \leq \bold{f}(\bold{p}) \leq \bold{b}$. 
Thus, there is a unique fixed point and the power updating function in (\ref{power updating function}) converges~to~it.
\end{proof}

\subsection{Feasibility Analysis}
In this section, we provide conditions to check the feasibility of the system. There is a one-to-one relation between a transmit power vector $\bold{p}=[p_1,p_2,\cdots,p_K]^\textnormal{T}$ and the SINR vector $\bm{\gamma}=[\gamma_1,\gamma_2,\cdots,\gamma_K]^\textnormal{T}$ \cite{Rasti2011}, that is:
\begin{equation}
\label{one to one relation}
p_i=\frac{\gamma_i}{h_i(\gamma_i+1)} \times \frac{\sigma^2}{1-\sum_{k=1}^{K}{\frac{\gamma_k}{\gamma_k+1}}}, \qquad  \forall i \in \mathcal{K}.
\end{equation}

\begin{definition}
The target SINR vector is feasible if a power vector $0 \leq \bold{p} \leq \overline{\bold{p}}$ exists that satisfies target SINRs of UEs, where it implies $0 \leq p_i \leq \overline{p}_i,~~\forall i \in \mathcal{K}$. Also, the system is feasible if the target SINR vector for all UEs is feasible, otherwise, the system is called infeasible.
\end{definition}

Given the target SINR $\widehat{\gamma}_i$ for each UE $i$ and by using (\ref{one to one relation}), we conclude that the target SINR vector is feasible if:
\begin{equation}
\label{feasible target SINR}
0 \leq \frac{\widehat{\gamma}_i}{h_i(\widehat{\gamma}_i+1)} \times \frac{\sigma^2}{1-\sum_{k=1}^{K}{\frac{\widehat{\gamma}_k}{\widehat{\gamma}_k+1}}} \leq \overline{p}_i, \qquad  \forall i \in \mathcal{K}.
\end{equation}
The intersection of feasible ranges for all $\widehat{\gamma}_i$ as a result of lower and upper bounds of (\ref{feasible target SINR}) indicates the feasibility condition.

\section{Numerical Results}
We consider a single square cell system with radius $100$~m where the BS is located at the midpoint and $5$ UEs are placed uniformly within the cell. 
The channel gains~are assumed to be generated using the path loss model $\textnormal{PL}_i(d)=\textnormal{PL}_0 + 10 \theta \log_{10}{d_i}$, in which $\textnormal{PL}_0$ is the constant path loss coefficient, $\theta$ is the path loss exponent, and $d_i$ is the distance between the $i$th UE and BS \cite{Tang}.
The power consumed by the circuit of BS and UEs are set as $P_\textnormal{BS}^\textnormal{C} = 30$ dBm and $P_i^\textnormal{C} = 20$ dBm, respectively (similar to \cite{Aslani2019} and \cite{Ng}). We set the power amplifier inefficiency of the UEs as  $\mu_i = 5$ (as in \cite{Tang}). The upper bound on the transmit power for all UEs is set as $\overline{p}_i=23$ dBm, similar to \cite{Tang} and \cite{Ng}.
The AWGN power at BS receiver is set as $\sigma^2=-113$ dBm, similar to \cite{DTPC} and \cite{Aslani}. 
According to (\ref{feasible target SINR}), the system is feasible for $\widehat{\gamma}_i \leq -7$~dB, $~\forall i \in \mathcal{K}$. 
Thus, we consider $\widehat{\gamma}_i$ belong to this feasibility range in all simulation scenarios.
For each $\widehat{\gamma}_i$, the value of $\delta_i$ is obtained by solving problem (\ref{problem delta}) using the scheme proposed in \cite{Aslani2019}.
We obtain all the numerical results by averaging over $1000$ independent snapshots with randomly generated location of UEs.
 
\begin{figure}
\centering
\includegraphics[width=2.5in]{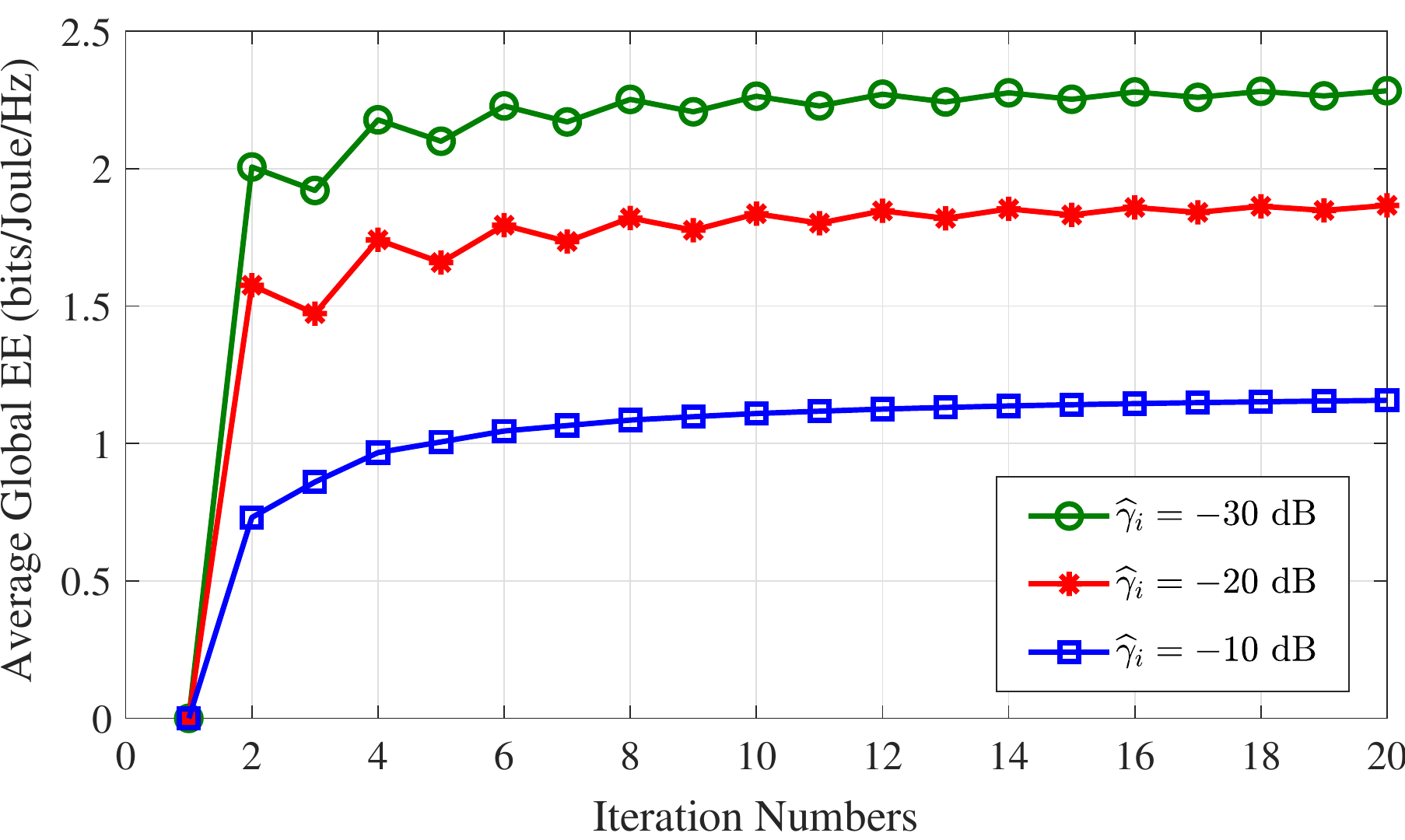}
\caption{Average GEE (bits/Joule/Hz) vs. iteration~number~for~different~$\widehat{\gamma}_i$.} 
\label{fig:convergence}
\end{figure}

We first investigate the convergence behavior of our iterative algorithm.
Fig. \ref{fig:convergence} shows average GEE vs. the iterations number for different values of $\widehat{\gamma}_i$. 
As seen in Fig \ref{fig:convergence}, average GEE for different values of $\widehat{\gamma}_i$ converges to the fixed point after $14$ iterations.
Thus, the performance of our proposed scheme after $14$ iterations is shown in the following case studies.
Another significant observation from Fig. \ref{fig:convergence} is that increasing $\widehat{\gamma}_i$ leads to a decreased GEE. To gain insight into, we examine the effect of the target SINR threshold on the GEE as follows.

\begin{figure}
\centering
\includegraphics[width=2.5in]{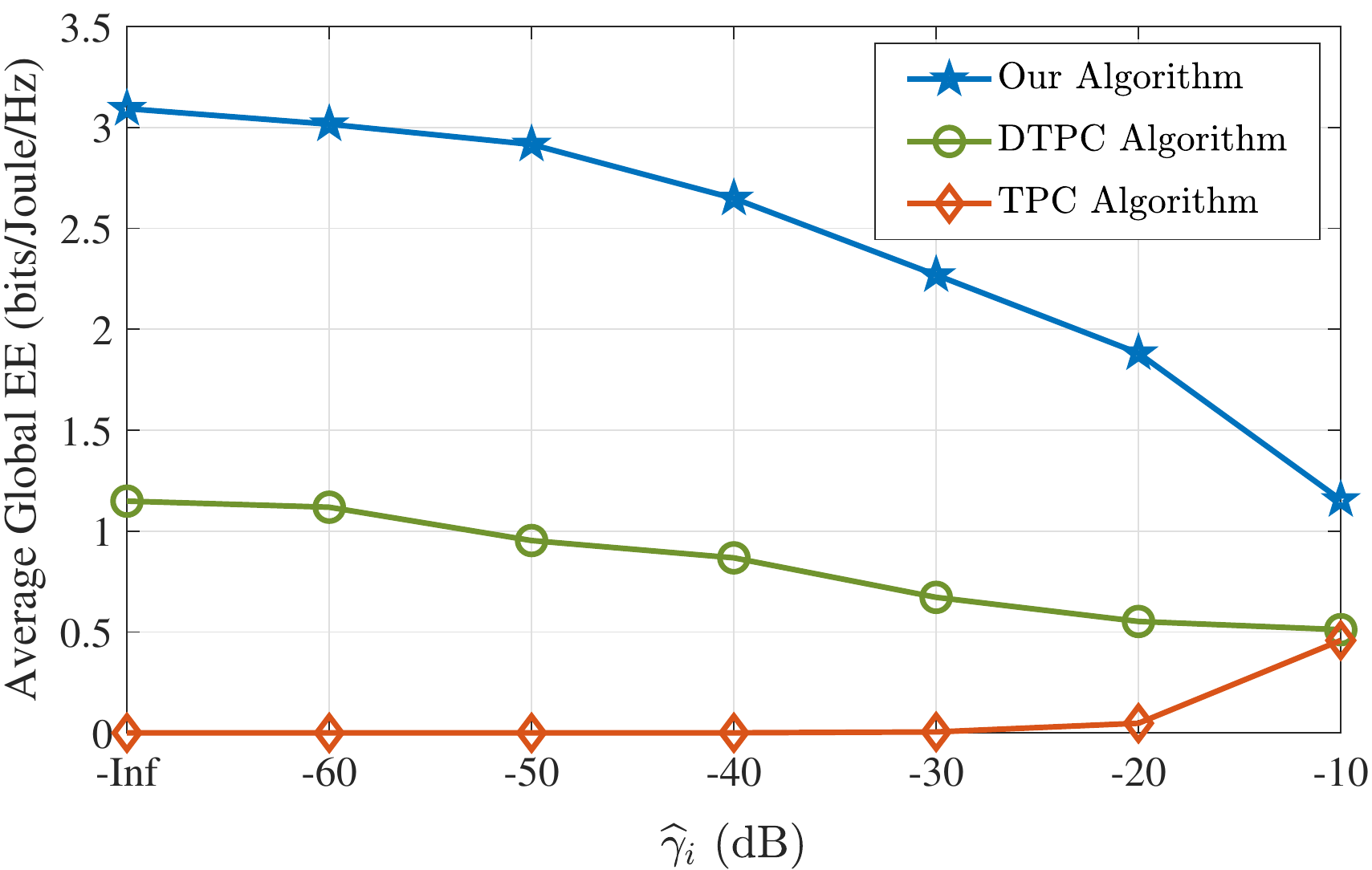}
\caption{Average GEE (bits/Joule/Hz) vs. $\widehat{\gamma}_i$ for different algorithms.} 
\label{fig:SINR}
\end{figure}

Fig. \ref{fig:SINR} shows average GEE vs. the value of $\widehat{\gamma}_i$ for different algorithms. As we observe from this figure, the GEE obtained by our proposed algorithm is a monotonically non-increasing function of the target SINR threshold. Indeed, the GEE reaches the highest value when $\widehat{\gamma}_i=$~-inf~dB ($\widehat{\gamma}_i=0$). The reason is that as $\widehat{\gamma}_i$ is high, the UEs have to transmit with a higher power in all channels to satisfy the SINR constraints leading to higher total power consumption. Also, the higher transmit power by all UEs including UEs with low channel gain results in higher interference which leads to lower total throughput. 
In contrast, when $\widehat{\gamma}_i$ is low, the UEs with high channel gain transmit with high power to increase the total throughput and UEs with low channel gains transmit with low power to satisfy the low SINR threshold which results in higher total throughput and lower total power consumption leading to higher GEE. This result has also been made in \cite{Aslani2019} for a different system model and resource allocation problem. 
Another significant observation from Fig. \ref{fig:SINR} is a demonstration of the behavior of our proposed algorithm compared to TPC \cite{TPC} and DTPC \cite{DTPC} algorithms.
From this figure, we see that our proposed scheme outperforms TPC and DTPC algorithms. More specifically, in the TPC algorithm, all UEs transmit in low power to satisfy the target SINR threshold which leads to lower total throughput and subsequently lower GEE. On the other hand, in the DTPC algorithm, some UEs transmit with high power to increase their achieved SINR which results in higher total power consumption and subsequently lower GEE. However, by controlling the transmit power through our proposed scheme, a balance between total throughput and total power consumption is achieved resulting in higher GEE.
 
\begin{figure}
\centering
\includegraphics[width=2.6in]{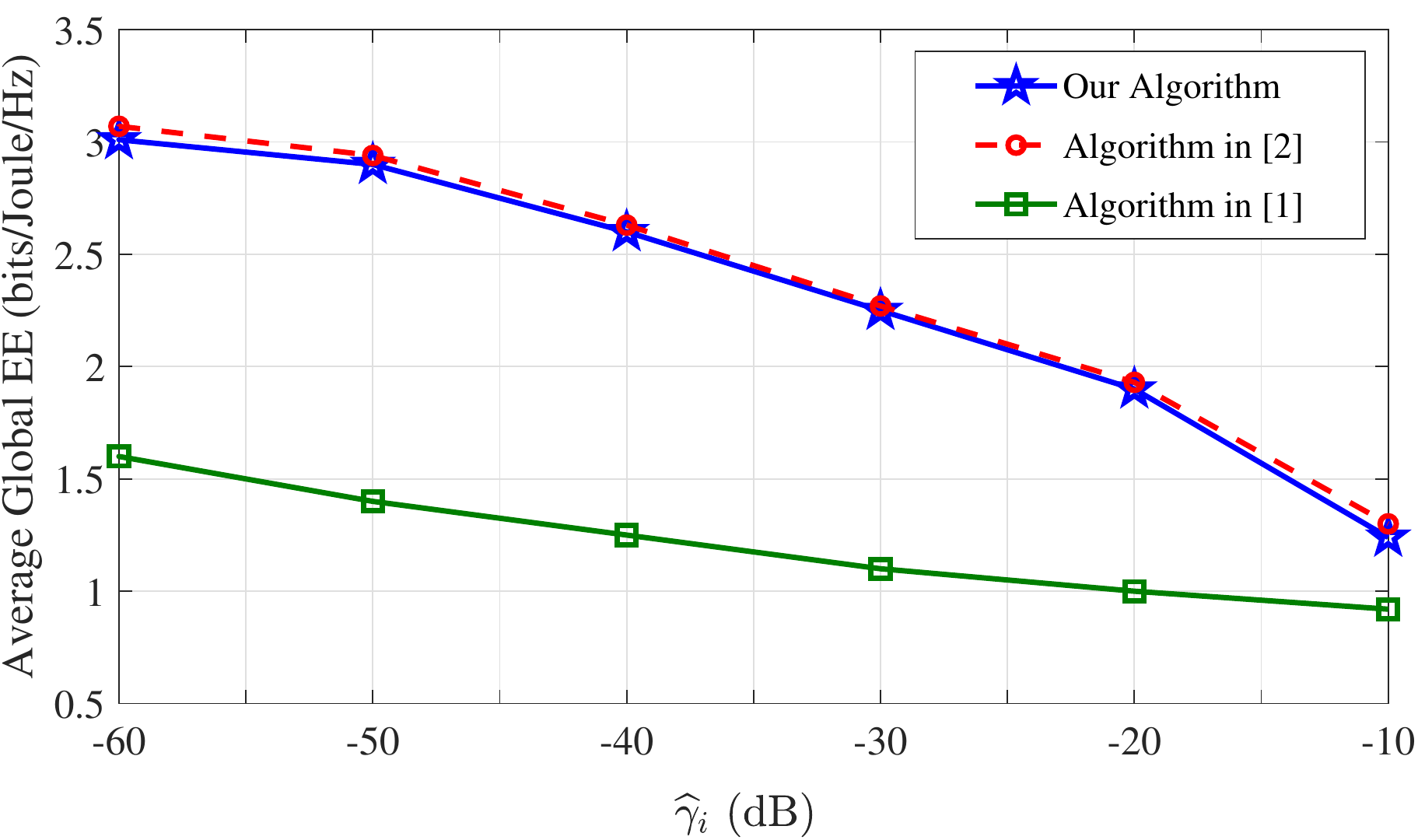}
\caption{Average GEE (bits/Joule/Hz) vs. $\widehat{\gamma}_i$ for different algorithms.}
\label{fig:compare}
\end{figure}

Finally, we provide the performance comparison between our proposed algorithm and other existing schemes in the literature which aim to maximize the EE. We consider the centralized scheme proposed in \cite{Aslani2019} to maximize the GEE, and the distributed scheme proposed in \cite{Zappone} to maximize the sum-EE. 
Fig. \ref{fig:compare} illustrates average GEE vs. the value of $\widehat{\gamma}_i$ for our proposed algorithm, the algorithm in \cite{Aslani2019}, and the algorithm in \cite{Zappone}. 
We observe that for all values of $\widehat{\gamma}_i$, our proposed distributed algorithm performs like the centralized algorithm in \cite{Aslani2019}. In the algorithm in \cite{Aslani2019} executed in the BS, the BS has the global CSI of all UEs and obtains the transmit power control policy in a centralized manner. However, in our algorithm executed in the UEs, each UE controls its transmit power in a distributed manner using its CSI only.
Additionally, our algorithm outperforms the algorithm in \cite{Zappone} for all values of $\widehat{\gamma}_i$. The reason is that in our proposed algorithm, each UE controls its transmit power to maximize the GEE and the interference is managed accordingly. However, the goal of the algorithm proposed in \cite{Zappone} is to maximize the summation of the UEs' EE, so each UE targets at maximizing its EE. Therefore, the transmit powers of UEs are not necessarily in accordance with the GEE maximization, and so the interference among the UEs is not managed to maximize GEE. This causes higher inter-UEs interference which results in higher total power consumption and lower total throughput leading to lower GEE in the algorithm in \cite{Zappone}.

\section{Conclusion}
In this paper, we studied the problem of GEE maximization subject to the minimum SINR of UEs in cellular networks. Since the stated problem was non-convex, we reformulated it as a MOO problem with the aim of minimizing the total power consumption and maximizing the total throughput, simultaneously, while satisfying UEs' target SINR. We employed the $\epsilon$-constraint method to address the problem. Finally, we proposed an iterative algorithm to control the transmit power of UEs in a distributed manner. We proved that our proposed iterative algorithm converges to its corresponding unique fixed point. Numerical results demonstrated that our proposed scheme rapidly converges and outperforms other algorithms. Simulation results also show that our distributed algorithm which obtains the maximum GEE using the local CSI performs like the centralized algorithm which maximizes the GEE by employing the global CSI.

\bibliographystyle{IEEEtran}
\bibliography{Mybib}

\end{document}